\documentclass[aps,prl,showpacs,superscriptaddress,preprintnumbers,floatfix]
{revtex4}
\usepackage{amsmath,amssymb}
\usepackage[dvips]{graphics}
\usepackage{epsfig}
\usepackage{psfrag}

\newcommand{\be}{\begin{equation}}
\newcommand{\ee}{\end{equation}}
\def\ba{\begin{array}}
\def\ea{\end{array}}

\font\psyra=psyr at 9pt
\def\tauup{\mbox{\psyra t}}
\newcommand{\bea}{\begin{eqnarray}}
\newcommand{\eea}{\end{eqnarray}}
\begin{document}

\title{Stochastic Quantization of the Ho\v{r}ava Gravity}

\author{Fu-Wen Shu}
\email{shufw@cqupt.edu.cn} \affiliation{College of Mathematics and
Physics, Chongqing University of Posts and Telecommunications,
Chongqing, 400065, China}
\author{Yong-Shi Wu}
\email{wu@physics.utah.edu} \affiliation{Department of Physics and
Astronomy, University of Utah, Salt Lake City, UT 84112, USA}

\date{\today}

\begin{abstract}

The stochastic quantization method is applied to the recent proposal
by Ho\v{r}ava for gravity. We show that in contrast to General
Relativity, the Ho\v{r}ava's action, satisfying the detailed balance
condition, has a stable, non-perturbative quantum vacuum when the
DeWitt parameter $\lambda$ is not greater than $1/3$, providing a
possible candidate for consistent quantum gravity.
\end{abstract}


\maketitle

{\sl Introduction.} The goal of formulating a consistent and
renormalizable quantum theory of gravity has been pursued for more
than half century. Attempts of quantizing General Relativity
(Einstein's theory of gravitation) have met tremendous difficulties.
On one hand, the canonical quantization is shown to be
perturbatively non-renormalizable in four
dimensions\cite{thooft,weinberg} and, therefore, loses
predictability, because the Einstein-Hilbert action is
non-polynomial. On the other hand, the Euclidean path integral
approach suffers the indefiniteness problem\cite{hawking}: Namely
the Einstein-Hilbert action is not positive-definite, because
conformal transformations can make the action arbitrarily negative.

A recent effort attempting to overcome these difficulties is the
proposal made by Ho\v{r}ava\cite{horava09}. (For the ideas that led
to this proposal, see also refs. \cite{horava1,horava2}.) This
proposal is a non-Lorentz invariant theory of gravity in 3+1
dimensions, inspired by the Lifshitz model\cite{lifshitz} studied in
condensed matter physics. At the microscopic (ultraviolet) level
this model exhibits anisotropic scaling between space and time, with
the dynamical critical exponent $z$ set equal to 3. (Namely, the
action is invariant under the scaling $x^i \rightarrow bx^i
(i=1,2,3), t\rightarrow b^z t$, where $z\neq 1$ violates the Lorentz
symmetry.) The action is assumed to satisfy the so-called detailed
balance condition, and is renormalizable by power counting. It is
argued that the renormalization group flow in the model approaches
an infrared (IR) fixed point theory with $z=1$, thus Einstein's
General Relativity (with local Lorentz symmetry) is naturally
emergent or recovered at the macroscopic level. It is this
perspective that has enabled the proposal to attract a lot of
interests in recent literature. Many papers have appeared to study
the classical solutions or consequences of the Ho\v{r}ava's proposal
(e.g. see refs. \cite{sm,calcagni,kiritsis}). A number of
fundamental questions remained unanswered. In this letter we report
a study of the most fundamental questions on Ho\v{r}ava gravity:
whether the action can really be quantized in a consistent and
non-perturbative manner? If yes, whether this will put any
constraint(s) on the parameters appearing in the action or not? (A
recent paper\cite{orlando} on the renormalizability of the model did
not address these issues, assuming no problem with quantization.)

Among the three existing -- canonical, path integral and stochastic
-- quantization approaches, only the last (stochastic quantization)
is constructive through stochastic differential equation, so that
the question of whether a stable vacuum (ground state) really exists
or not can be easily investigated and answered. Also it has the
great advantage\cite{wu} of no need for gauge-fixing when applied to
theories with gauge symmetry. In this letter we apply stochastic
quantization to the Ho\v{r}ava gravity, where the gauge symmetry is
spatial diffeomorphisms. We will show that the quantized theory with
a stable vacuum indeed exists only when the parameter $\lambda$ in
the DeWitt metric in the space of three-metrics is not greater than
a critical value 1/3: (i.e. $\lambda\leq \lambda_c=1/3$). This is
the range of the values of $\lambda$ for which Ho\v{r}ava's action
may make sense for a consistent quantum theory of gravity. (In
contrast, stochastic quantization of General Relativity does not
lead to a stable vacuum (ground) state. See below.)

{\sl Preliminaries.} Assume the spacetime allows a
$(3+1)$-decomposition:
\be ds_4^2= - N^2  dt^2 + g_{ij} (dx^i - N^i dt) (dx^j - N^j dt)\,,
\ee
where $g_{ij} (i,j=1,2,3)$ is the 3-metric, $N$ and $N_i$ are the
lapse function and shift vector, respectively. The Ho\v{r}ava
action with $z=3$ is given by\cite{horava09}
\begin{equation}\label{gaction}
S=\int dt d^3x
\sqrt{g}N\left[\frac{2}{\kappa^2}K_{ij}\mathcal{G}^{ijkl}
K_{kl}+\frac{\kappa^2}{8}E^{ij}\mathcal{G}_{ijkl}E^{kl}\right],
\end{equation}
where $g$ denotes the determinant of the 3-metric $g_{ij}$ and
$\kappa^2$ is the coupling constant, to be identified with $32\pi
Gc$ in the IR regime with $z=1$ ($G$ and $c$ the Newton's
gravitational constant and the speed of light, respectively). The
extrinsic curvature $K_{ij}$ and the DeWitt metric
$\mathcal{G}^{ijkl}$ in \eqref{gaction} are defined by
\bea K_{ij}& = &\frac{1}{2N} (\dot g_{ij} - \nabla_i N_j - \nabla_j
N_i),\\
\mathcal{G}^{ijkl}&=&\frac12\left(g^{ik}g^{jl}+g^{il}g^{jk}\right)
-\lambda g^{ij}g^{kl} \eea
with $\lambda$ a free parameter. The potential term in
\eqref{gaction}, when $E^{ij}$ is given by $
\sqrt{g}E^{ij}=\frac{\delta W}{\delta g_{ij}},$ is said to satisfy
the so-called detailed balance condition. Ho\v{r}ava took $W$ to
be
\be W=\frac{1}{w^2}\int \omega_3(\Gamma)+\mu \int d^3x\sqrt{g}
(R-2\Lambda_W).\label{potential}\ee
Here $\mu\,,w$ and $\Lambda_W$ are coupling constants, and
$\omega_3$ is the gravitational Chern-Simons term:
\be \omega_3\equiv \mbox{Tr}\left(\Gamma\wedge
d\Gamma+\frac23\Gamma\wedge\Gamma\wedge\Gamma\right),\label{CS} \ee
with $\Gamma$ the Christoffel symbols. Simple dimensional analysis
for the coupling constants shows that the theory is ultraviolet (UV)
renormalizable\cite{calcagni}. The renormalizability beyond the
power counting of this theory has recently been confirmed in
\cite{orlando}, assuming no problem with quantization. Here we will
examine the more fundamental question of the non-perturbative
existence of quantum vacuum.\\

{\sl Stochastic Quantization.} Stochastic quantization\cite{wu}
has been proved to be an effective tool for quantizing a field
theory, in particular a gauge theory\cite{huffel,namiki}.
Stochastic quantization is based on the principle that quantum
dynamics of a $d$-dimensional system is equivalent to classical
equilibrium statistical mechanics of a $(d+1)$-dimensional system.
The essence of stochastic quantization is to use a stochastic
evolution -- the Langevin equation -- in fictitious time, driven
by white noises, to construct the equilibrium state corresponding
to the quantum ground state. The existence of an equilibrium state
can be proved or disproved by studying the corresponding
Fokker-Planck equation associated with the Langevin equation.In
this spirit, we start with the Langevin equation of the Ho\v{r}ava
gravity:
\bea
\begin{cases}
\dot{N}=-\frac{1}{\sqrt{g}}\frac{\delta S_E}{\delta N}+\eta,\\
\dot{N_i}=-\frac{1}{\sqrt{g}}\frac{\delta S_E}{\delta N^i}+\zeta_i,\\
\dot{g}^{I}=-\mathcal{G}^{IJ}\partial_{J}S_{E}+\xi^I,
\end{cases}\label{langevin} \eea
where the dot represents derivative with respect to the fictitious
time $\tau$ and following notations have been introduced:
$$
g_{ij}\equiv g^I,\ \ \ \ \mathcal{G}^{IJ}\equiv \mathcal{G}_{ijkl},
\ \ \ \ \partial_I S_{E}\equiv \frac1{\sqrt{g}}\frac{\delta
S_{E}}{\delta g_{ij}}.
$$
In eq. \eqref{langevin}, $\eta^I$, $\zeta_i$ and $\xi^I$ are noises,
and $S_{E}$ is the Euclidean version of the action \eqref{gaction}.

Note that the indices $I$, $J$ (=1,2,\ldots 6) are raised and
lowered by $\mathcal{G}^{IJ}$ and its inverse $\mathcal{G}_{IJ}$.
The stochastic correlation of a gauge invariant functional
$\mathcal{F}(N,N_i,g_I)$ is defined as the expectation value of
the functional with respect to the noises
\be <\mathcal{F}(N,N_i,g_I)>\sim \int
\mathcal{D}[\eta]\mathcal{D}[\zeta]\mathcal{D}[\xi]
\mathcal{F}(N,N_i,g_I)\exp\left[-\frac14\int d\tauup d^3xd\tau
\sqrt{g}N(\eta^2+g^{ij}\zeta_i\zeta_j+\mathcal{G}^{IJ}\xi_I\xi_J)\right]
\label{correlation}
\ee 
where $g^{ij}$ and $\mathcal{G}^{IJ}$ are solutions of the Langevin
equation \eqref{langevin} and hence are functions of $\zeta^i$ and
$\xi^I$, respectively. The Wick rotation to imaginary time $\tauup$
has been applied and $\tau$ is the fictitious time. Eq.
(\ref{correlation}) indicates that the noises $\zeta_i$ and $\xi_I$
are not Gaussian. As suggested in \cite{orlando}, one can overcome
this difficulty by introducing a set of new noises via vielbein.
That is $\zeta^a\equiv e^a{}_i\zeta^i$, $\xi^A\equiv E^A{}_I\xi^I, $
and its inverse $ \zeta^i= e^i{}_a\zeta^a$, $\xi^I= E^I{}_A\xi^A, $
where $e^a{}_i$ and $E_A{}^I$ are the vielbeins. The following
relations hold
\bea e_a{}^ie_b{}^jg_{ij}=\delta_{ab},\ \ \
E_A{}^IE_B{}^J\mathcal{G}_{IJ}=\delta_{AB},\\
e_a{}^ie_b{}^j\delta^{ab}=g^{ij},\ \ \
E_A{}^IE_B{}^J\delta^{AB}=\mathcal{G}^{IJ}. \eea
The new noises turn out to be Gaussian and we have
\bea
&&<\eta(x,\tau)>=0,\ \ <\zeta^a(x,\tau)>=0,\ \ <\xi^A(x,\tau)>=0,\\
&&<\eta(x,\tau)\eta(y,\tau')>=2\delta(x-y)\delta(\tau-\tau'), \\
&&<\zeta^a(x,\tau)\zeta^b(y,\tau')>=2\delta^{ab}\delta(x-y)\delta(\tau-\tau'), \\
&&<\xi^A(x,\tau)\xi^B(y,\tau')>=2\delta^{AB}\delta(x-y)\delta(\tau-\tau').
\eea
(Here $x$ stands for Euclidean coordinates $(x^i,\tauup)$.) The
Langevin equation \eqref{langevin} then becomes
\bea
\begin{cases}
\dot{N}=-\frac{1}{\sqrt{g}}\frac{\delta S_E}{\delta N}+\eta,\\
\dot{N_i}=-\frac{1}{\sqrt{g}}\frac{\delta S_E}{\delta N^i}+\zeta_ae^a{}_i,\\
\dot{g}^{I}=-\mathcal{G}^{IJ}\partial_{J}S_{E}+\xi^AE_A{}^I,
\end{cases}\label{langevin1} \eea
and the correlation functional is redefined with respect to $\eta$,
$\zeta^a$ and $\xi^A$ by
\be <\mathcal{F}(N,N_i,g_I)>\sim \int
\mathcal{D}[\eta]\mathcal{D}[\zeta]\mathcal{D}[\xi]
\mathcal{F}(N,N_i,g_I)\exp\left[-\frac14\int d\tauup
d^3xd\tau\sqrt{g}N(\eta^2+\zeta^a\zeta_a+\xi^A\xi_A)\right],
\label{correlation1}
\ee 
which is obviously Gaussian as desired.

To study whether the Langevin process \eqref{langevin1} really
converges to a stationary equilibrium distribution, we examine the
probability density functional associated with it:
\be P(N,N^i,g_I,\tau)=\frac{\exp\left[-\frac14\int d\tauup
d^3xd\tau\sqrt{g}N(\eta^2+\zeta^a\zeta_a+\xi^A\xi_A)\right]}
{\int\mathcal{D}[\eta]\mathcal{D}[\zeta]\mathcal{D}[\xi]\exp\left[-\frac14\int
d\tauup
d^3xd\tau\sqrt{g}N(\eta^2+\zeta^a\zeta_a+\xi^A\xi_A)\right]}.\label{prob}
\ee
We introduce
\be Q(N,N^i,g_I,\tau) \equiv
P(N,N^i,g_I,\tau)e^{S_E/2}.\label{definition}
\ee 
and the Fokker-Planck equation for the probability distribution is
\bea \frac{\partial Q(N,N^i,g_I,\tau)}{\partial
\tau}=-\mathcal{H}_{FP}Q(N,N^i,g_I,\tau), \label{fp2}
\eea 
where the Fokker-Planck Hamiltonian $\mathcal{H}_{FP}$ is of the
form
\bea \mathcal{H}_{FP}=a^{\dagger}a+g^{ij}a_i{}^{\dagger}a_j
+\mathcal{G}^{IJ}\mathcal{A}_I{}^{\dagger}\mathcal{A}_J.
\label{hamiltonian} \eea
Here
$$
a=i\pi+\frac12\frac1{\sqrt{g}}\frac{\delta S_E}{\delta N},\ \
a^i=i\pi^i+\frac12\frac1{\sqrt{g}}\frac{\delta S_E}{\delta N_i},\ \
\mathcal{A}^I=i \pi^I+\frac12\partial^IS_{E},
$$
with $\pi$, $\pi^i$ and $\pi^I$, respectively, the conjugate
momenta of $N$, $N^i$ and $g^I$:
$\pi=-i\frac1{\sqrt{g}}\frac{\delta}{\delta N}$,
$\pi^i=-i\frac1{\sqrt{g}}\frac{\delta}{\delta N_i}$, $\pi_I=-i
\partial_I$.
The time independent eigenvalue equation associated with Eq.
\eqref{fp2} is
\bea \mathcal{H}_{FP} Q_n(N,N^i,g_I,\tau)=E_n Q_n(N,N^i,g_I,\tau).
\eea
The solutions of Eq. \eqref{fp2} lead to the general solution
\bea
P(N,N^i,g_I,\tau)=\sum_{n=0}^{\infty}a_nQ_n(N,N^i,g_I,\tau)
e^{-S_{E}/2-E_n\tau}.\label{prob2}
\eea 
The stationary candidate for the equilibrium state is given by
$Q_0(N,N^i,g_I)=e^{-S_{E}/2}$ with $E_0=0$. From the above formula
we see that the theory will approach an equilibrium state for large
$\tau$ if and only if all other $E_n>0$. To find the condition(s)
under which the Fokker-Planck Hamiltonian \eqref{hamiltonian} is
non-negative definite, we note that the sum of the first two terms
$(a^{\dagger}a+g^{ij}a_i{}^{\dagger}a_j)$ always respects this
property. So we only need to find condition(s) under which the
eigenvalues of the DeWitt metric $\mathcal{G}^{IJ}$ are all
non-negative. By a straightforward computation, the desired
condition is found to be $\lambda \leq 1/3$: When $\lambda<1/3$,
$\mathcal{G}^{IJ}$ is positive definite; if $\lambda>1/3$, one and
only one eigenvalue of $\mathcal{G}^{IJ}$ becomes negative. At the
critical value $\lambda=1/3$, one eigenvalue of $\mathcal{G}^{IJ}$
becomes zero, while all others remain positive.

Thus the Fokker-Planck Hamiltonian is non-negative definite if
$\lambda\le 1/3$, and the theory approaches an equilibrium in this
case: It follows from eq. \eqref{prob2} that
\bea \lim_{\tau\rightarrow \infty} P(N,N^i,g_I,\tau)=a_0 e^{-S_{E}},
\label{prob1}
\eea 
where $a_0$ is determined by the normalization condition. Note that
this result is independent of the initial conditions. Any equal-time
correlation function \eqref{correlation1}, if invariant under
spatial diffeomorphisms, tends to its equilibrium value for large
time $\tau$. Therefore, though the solution given by \eqref{prob1}
is always a stationary state for the Fokker-Planck equation, it
represents an equilibrium state (or a stable ground state) reached
at large time $\tau$ only when $\lambda\le 1/3$. In contrast, a
similar result would not be obtained with stochastic quantization of
Einstein's gravity, which corresponds to $\lambda=1>1/3$, since the
associated Fokker-Planck Hamiltonian is not positive definite and
hence does not lead to an equilibrium state at large fictitious
times.

In the above derivation, the detailed balance condition is crucial
for the Ho\v{r}ava gravity to have a stable vacuum when
$\lambda<1/3$. In fact, with the detailed balance condition
satisfied at short distances, $S_{E}$ is of the form
$$
S_{E}=\int \mathcal{G}^{IJ}(K_IK_J+\alpha E_IE_J),
$$
where $\alpha >0$ and $E_I=\partial_I W$ with $W$ given by
\eqref{potential}. $S_E$ has a similar structure to eq.
\eqref{hamiltonian}, so it is positive definite for $\lambda<1/3$
and indefinite for $\lambda>1/3$. As a consequence, the state
\eqref{prob1} is a physical ground state for $\lambda<1/3$ and is
unstable for $\lambda>1/3$.

We have seen that $\lambda_c=1/3$ is a critical value for the
theory: Above it the quantized theory does not make sense, while
the opposite is true below it. Exactly at $\lambda=\lambda_c$,
extra zero modes develop for the DeWitt metric $\mathcal{G}^{IJ}$
and, hence, for eq. \eqref{fp2} as well. This implies that the
gauge symmetry of the theory is enhanced, which now includes local
Weyl transformations as already observed in ref. \cite{horava09}.
It would be extremely interesting to understand the fate of the
enhanced gauge symmetry in the quantized theory. Anyway, in
principle stochastic quantization method should be applicable at
$\lambda=1/3$, and the appearance of extra zero modes does not
destroy the stability of the new vacuum, though there are subtle
issues to be resolved.

{\sl Conclusions and Discussions.} In summary, we have applied
stochastic quantization to the Ho\v{r}ava gravity. By analyzing
the associated Fokker-Planck equation, we have found that with
$\lambda<1/3$ the system will approach to equilibrium as the
fictitious time goes to infinity, giving rise to a stable vacuum
state for the quantized theory. The key to this property is the
detailed balance condition obeyed by the Ho\v{r}ava action. When
$\lambda>1/3$, stochastic quantization does not make sense because
of development of a negative mode. The $\lambda=1/3$ case would be
probably alright, but needs more careful examination.

In ref. \cite{horava09}, to make sense of the speed of light in
the IR regime with $z=1$, one needs $\Lambda_W/(1-3\lambda)$ to be
positive. Our constraint $\lambda<1/3$ for the stability of
gravity vacuum further constrains the cosmological constant to be
positive: $\Lambda_W >0$. This agrees with cosmological
observations\cite{wmap5}.

Our suggestion opens the door for using stochastic quantization to
numerically study the quantized Ho\v{r}ava gravity, in particular
to check whether the renormalization group would indeed change the
value of $z$ from $z=3$ in the UV regime to $z=1$ in the IR
regime.

Finally, it should be noted that the stochastic quantization applied
in this letter is the standard one that introduces a fictitious
time. This is different from the one used in ref. \cite{orlando},
where the time for stochastic evolution is identified with the real
time. In this reference, for the purpose of studying the
renormalizability of Ho\v{r}ava gravity, they have explored the fact
that like any Lifshitz-type models, the Ho\v{r}ava gravity can be
viewed as stochastic quantization of a lower dimensional
theory\cite{dijkgraaf}, which in the present case is
three-dimensional topological massive gravity.\\

\noindent {\sl Acknowledgments.} This work is partially supported
by a grant from FQXi. One of us (F.W.) thanks Department of
Physics and Astronomy, University of Utah for warm hospitality,
where this work was done. F.W. is supported by a grant from CQUPT.
YSW is supported by US NSF grant PHY-0756958.

\end{document}